\documentstyle[12pt]{article}
\input{epsf}

\newcommand{\beq}{\begin{eqnarray}}
\newcommand{\eeq}{\end{eqnarray}}

\newcommand{\cket}[1]{| #1 \rangle}
\newcommand{\bra}[1]{\langle #1 |}

\newcommand{\pslash}{{p\hspace{-5pt}/}}

\newcommand{\gpNN}{g_{\pi NN}}
\newcommand{\geNN}{g_{\eta NN}}
\newcommand{\gpNR}{g_{\pi NN^*}}
\newcommand{\geNR}{g_{\eta NN^*}}

\begin{document}

\begin{center}
{\large Suppression of $\pi NN(1535)$ Coupling
in the QCD Sum Rule \\ }
\vspace*{0.5cm}
 { D. Jido\footnote{e-mail address: jido@th.phys.titech.ac.jp} and M. Oka \\
 Department of Physics, Tokyo Institute of Technology\\
 Meguro, Tokyo 152  Japan}

\vspace*{0.5cm}
{A. Hosaka \\
Numazu College of Technology \\
3600 Ooka, Numazu 410 Japan}
\end{center}
\vspace*{1cm}
\abstract{
The $\pi NN^*$ and $\eta NN^*$ coupling constants are studied
based on the two point functions between the
vacuum and a one meson state in the soft meson limit.  
In terms of a suitably constructed interpolating field for the nucleon 
and resonances, 
we have found that the $\pi NN^*$ couplings vanish, while $\eta NN^*$ 
couplings remain finite.  
This result explains a relatively suppressed coupling of $\pi NN(1535)$ 
as compared with others.  
We compare the present result with predictions of low energy
effective models.
}

\newpage

Baryon resonances provide a good testing ground of
effective models for QCD at low energy.
Not only the resonance masses but also transition matrix elements are
subject to recent research interest.
New data for these quantities will soon become available
from facilities such as TJNL (former CEBAF), ELSA, MAMI
and possibly Spring8~\cite{INT}.
In particular, transition matrix elements are very sensitive to details
of resonance wave functions, and therefore, they will provide strong
constraints on the low energy effective models.

Among various baryon resonances, negative parity resonances $N^*$
have particularly interesting properties.
For example, $N(1535)$  has relatively large branching 
ratio for the decay $N(1535) \to \eta N$ which is comparative to that 
of $N(1535) \to \pi N$.  
Considering the difference in the available phase spaces,
this leads to a relatively
large coupling constant for $\eta NN(1535)$.
One may also look at the problem in the following way.  
Using the experimental decay widths of the resonance, we obtain
$\gpNR \sim 0.7$ and $\geNR \sim 2$.  
These values are in fact much smaller than those in the $NN$ sector:  
$\gpNN \sim 13$ and $\geNN \leq 5$.  
Furthermore, the pion couples weaker than the eta in the $NN^*$ sector, as 
opposed to the $NN$ sector.  
Thus, one may ask why the coupling $\gpNR$ is suppressed so much 
as compared with other couplings.  
We will look at the problem of the couplings 
from this point of view in this paper.

Previously, Jido, Kodama and Oka~\cite{jko}
have studied masses  of 
negative parity octet baryons  in the QCD sum
rule~\cite{svz,rry}.
A method to extract the information of $N^*$ from a 
correlation function has been formulated.
The resulting masses are generally in good agreement with data.  
One of the motivations in the previous work
was to study chiral properties of the negative parity baryons.
The behavior of the masses as functions of the quark condensate shows
that the negative parity baryons may be regarded as chiral
partners of the ground state baryons.
It is then very interesting to study the
coupling constants in the same framework.

Our starting point is the observation that an interpolating field
can couple to both positive and negative parity
baryons~\cite{cdks}.
The general interpolating field without derivatives for the nucleon $J_N$
is given by a superposition of two independent terms~\cite{ept}:
\begin{eqnarray}
   J_{N}(x;t) & = & \varepsilon^{abc} [(u_{a}(x)Cd_{b}(x))
   \gamma_{5} u_{c}(x) + t (u_{a}(x) C \gamma_{5} d_{b}(x))
   u_{c}(x)] \, ,
        \label{eq:nucur}
\end{eqnarray}
where  $a$, $b$ and $c$ are color indices, $C = i \gamma_{2} \gamma_{0}$
(in the standard notation) is the charge conjugation matrix, and
$t$ is the mixing parameter of the two
independent interpolating fields.
$J_{N}(x;t=-1)$, which is called the ``Ioffe's interpolating field''\cite{i},
is commonly used for the study of the ground state
nucleon in the QCD sum rule since it is almost
optimized for the lowest-lying nucleon~\cite{ept}.
For the study of $N^*$ we found in ref.~\cite{jko}
that the field $J_{N}(x;t=0.8)$ couples strongly to the negative-parity 
states.

The correlation function constructed from $J_{N}$, therefore, contains
information of both parity baryons.
In ref.~\cite{jko}, a method to separate the term of either
positive or negative parity state from the other exclusively has been
developed.
Since the correlation function consists of two terms with
opposite chirality,  their sum and difference have a definite parity.

In order to calculate the meson-baryon coupling constants, we follow the 
method
used by Shiomi and Hatsuda~\cite{sh}.
They studied the $\pi NN$ coupling constant $\gpNN$
by using the two point function between
the vacuum and a one meson state in the soft meson limit $(q^{\mu}
\to 0)$.
The relevant correlation function is
\beq
   \Pi^{m}(p) &=& i \int d^4 x \, e^{ip\cdot x}
        \bra{0} TJ_{N}(x) \bar{J_{N}}(0) \cket{m(q=0)} \nonumber \\
        &=&
        i \gamma_{5} (\Pi_{0}^{m}(p^{2})  + \Pi_{1}^{m}(p^{2}) \pslash ) \, ,
   \label{eq:cor}
\eeq
where $J_{N}$ is defined in (\ref{eq:nucur}), and $m$ denotes
either $\pi$ or $\eta$.
The parameter $t$ will be chosen suitably depending on whether $J_{N}$
should be coupled strongly to positive or negative parity baryons.

Let us first look at the phenomenological side of the correlation
function to see how information of the negative parity nucleon
can be extracted.
We shall see this for the $\pi NN^*$ coupling ($m = \pi$).
The phenomenological $\pi NN^*$
interaction lagrangian is defined by
\begin{equation}
    \label{eq:defpNN}
    {\cal L}_{\pi NN^*} = \gpNR \bar{N}^{*} \tau^{i} \pi^{i} N,
\end{equation}
where $N$ and $N^{*}$ are the field operators for the positive
and negative parity nucleons,
$\pi^{i}$ is the pion field, and $\tau^{i}(i =
1,2,3)$ are the Pauli matrices for isospin.
From the
Lagrangian (\ref{eq:defpNN}), the $\pi NN^{*}$ contribution in the
$\Pi^{\pi}(p)$ is given in the soft pion limit by
\begin{equation}
   \label{eq:pNN*}
    \gpNR \lambda_{N} \lambda_{N^{*}}
    \left[{p^{2} + m_{N}m_{N^{*}} \over (p^{2} - m_{N}^{2}) (p^{2} -
    m_{N^{*}}^2)} + { \pslash (m_{N} + m_{N^{*}}) \over (p^{2} -
    m_{N}^{2}) (p^{2} - m_{N^{*}}^2)}\right]i\gamma_{5} \, ,
\end{equation}
where $\lambda_{N}$ and $\lambda_{N^{*}}$ are defined by
$\langle 0 | J_{N} | N\rangle = \lambda_{N} u_{N}$ and 
$\bra{0} J_{N}\cket{N^{*}} = i \gamma_{5}
\lambda_{N^*} u_{N^*}$, respectively, with
$u_{N}$ and $u_{N^*}$ being the Dirac spinor for $N$ and $N^*$.  
We note that  there appear two terms in (\ref{eq:pNN*}); one
proportional to $\gamma_{5}$ and the other proportional to
$\pslash \gamma_{5}$.
In contrast, the $\pi NN$ contribution has only one term,
\begin{equation}
   \gpNN \lambda_{N}^{\;2} {i \gamma_{5} \over
    p^{2} - m_{N}^{\; 2}} \, ,    \label{eq:pheNN}
\end{equation}
as is derived from the $\pi NN$ interaction lagrangian
\begin{equation}
    {\cal L}_{\pi NN} = \gpNN \bar{N} i \gamma_{5} \tau^{i} \pi^{i}
    N\, .
\end{equation}
We note that (\ref{eq:pheNN}) is also obtained by replacing
$M_{N^*}$ by $-M_N$ in (\ref{eq:pNN*}).

In the soft pion limit,
Shiomi and Hatsuda studied the sum rule using the non-vanishing term
of (\ref{eq:pheNN}) and found that the resulting $\pi NN$ coupling constant
satisfies the
Goldberger-Treiman relation with $g_{A}=1$~\cite{sh}.
Recently, Birse and Krippa also studied the coupling constant
$\gpNR$ at a nonzero pion momentum~\cite{bk}.
In our case for the $\pi NN^*$ coupling constant,
we study the term proportional to $\pslash \gamma_{5}$,  
since that term exclusively contains transitions of 
$N^* \to \pi N$.  
There is a problem, however, that it contains not only the transitions 
from negative parity resonances but also those from positive parity 
resonances, the lowest of which is $N(1440)$.  
Such a term is, however, proportional to
the mass difference, e.g. $M_{N(1440)} - M_{N}$, unlike the sum as in
the second term of (\ref{eq:pNN*}).
Thus the contribution from positive parity resonances is expected to 
be relatively
suppressed as compared with that of negative parity resonances, at least in 
the low mass region.
Moreover, by choosing the
mixing parameter $t \sim 0.8$, 
the interpolating field (\ref{eq:nucur}) is made to
couple strongly to negative parity states, and 
the contribution from positive parity resonances is 
expected to be further suppressed.

The sum rule for the $\eta NN^{*}$ coupling is
similarly constructed
by replacing the isospin matrices $\tau$ in the $\pi NN^*$
coupling by the unit matrix.
For example, the interaction lagrangian for the $\eta NN^*$ coupling
is written as
\begin{equation}
    {\cal L}_{\eta NN^{*}} = \geNR \bar{N}^{*} \tau^{0} \eta N,
\end{equation}
where $\tau^{0} = 1$.

The correlation function is now computed by the operator product
expansion (OPE) perturbatively in the deep Euclidean region.
The result for the terms of $i \pslash \gamma_{5}$ takes the
following form
\begin{eqnarray}
    \Pi^{\rm OPE}(p) & = & i \int d^{4}x \, e^{ip\cdot x} \, \bra{0}{\rm T}
      J_{N}(x;s) \bar{J}_{N}(0;t) \cket{m}
      \label{eq:PiOPE}\\
    & \equiv & i \pslash \gamma_{5} \left[ {\cal C}_{4} \ln(-p^{2}) +
        {\cal C}_{6} {1 \over p^{2}} + {\cal C}_{8} {1 \over p^{4}} + \cdots
        \right] + i\gamma_{5} \left[ {\cal C}_{3} p^{2} \ln(-p^{2}) + \cdots
        \right], \nonumber
\end{eqnarray}
where we allow to use the different mixing parameters for the interpolating
fields such that $\bar{J}_N (0; t \sim 0.8)$ couples dominantly to the
$N^*$ state,
while $J_N(x; s = -1)$ to the $N$ state.
Note that the terms of $i \pslash \gamma_{5}$ are of even dimension.
The correlation function (\ref{eq:PiOPE}) has been calculated up to
dimension 8, ignoring higher order terms in $m_{q}$ and $\alpha_{s}$.
The results are
\begin{eqnarray}
{\cal C}_{4} & \sim &
        m_{q} \bra{0} \bar{q}i \gamma_{5} q \cket{m}
        \stackrel{m_q \to 0}{\longrightarrow} 0 \\
{\cal C}_{6} & = & -{s-t \over 4} \left[
    \langle \bar{d}d \rangle \bra{0} \bar{u} i \gamma_{5} u \cket{m} +
    \langle \bar{u}u \rangle \bra{0} \bar{d} i \gamma_{5} d \cket{m}
     \right] \\
{\cal C}_{8} & = & - {s-t \over 144} \left[
    25 (\langle \bar{d}gG \cdot \sigma d \rangle
        \bra{0} \bar{u} i \gamma_{5} u \cket{m} +
    \langle \bar{u}gG \cdot \sigma u \rangle
        \bra{0} \bar{d} i \gamma_{5} d \cket{m}) \right.  \nonumber \\
    & &  \left. - 7 ( \langle \bar{d}d \rangle
        \bra{0} \bar{u} i \gamma_{5} gG \cdot \sigma u \cket{m} +
     \langle \bar{u}u \rangle
        \bra{0} \bar{d} i \gamma_{5} gG \cdot \sigma d \cket{m}) \right]
\end{eqnarray}
where
$\langle \bar q q \rangle = \bra{0} \bar q q \cket{0}$
and
we assume the vacuum saturation for four-quark matrix elements.
For the calculation of the matrix element of the operators between the
vacuum and one meson state, we use the soft meson
theorem:
\begin{equation}
    \bra{0} {\cal O}(0) \cket{m^{i}(q)} \stackrel{q \rightarrow 0}
    {\longrightarrow} - {1 \over \sqrt{2} f_{m}} \int d^{3} x  \, \bra{0}
    [i J^{i}_{50}(x),{\cal O}(0)] \cket{0},
\end{equation}
where $J^{i}_{5\mu}(x)
= \bar{q}(x) \gamma_{\mu} \gamma_{5} (\lambda^{i} / 2) q(x)$
and $f_m$ is the decay constant of the meson $m$.
We apply this formula to both the pion and the eta, obtaining the
following relations:
\begin{eqnarray}
  \label{eq:uum}
    \bra{0} \bar{u}i \gamma_{5} u \cket{m} & = & -
        {\alpha_{m} \over f_{m}} \langle \bar{u}u \rangle \, , \\
  \label{eq:ddm}
    \bra{0} \bar{d}i \gamma_{5} d \cket{m} & = & \pm
        {\alpha_{m} \over f_{m}}  \langle \bar{d}d  \rangle \, ,  \\
  \label{eq:uGum}
    \bra{0} \bar{u}i \gamma_{5} G\cdot \sigma u\cket{m}
     & = & - {\alpha_{m} \over f_{m}}
         \langle \bar{u}G\cdot \sigma u \rangle  \, , \\
  \label{eq:dGdm}
    \bra{0} \bar{d}i \gamma_{5} G\cdot \sigma d \cket{m}
     & = & \pm {\alpha_{m} \over f_{m}}
         \langle \bar{d} G\cdot \sigma d \rangle \, ,
\end{eqnarray}
where $\alpha_{\pi} = 1/\sqrt{2}$ and $\alpha_{\eta} = 1/\sqrt{6}$.
Note that the sign change in (\ref{eq:ddm}) and (\ref{eq:dGdm})
is the only source for the difference between
the pion and eta matrix elements.
This originates from the different isospin structure:
$\pi^{0} \sim \frac{1}{\sqrt{2}}(\bar{u}u - \bar{d}d)$,
while $\eta \sim \eta_8
\sim \frac{1}{\sqrt{6}}(\bar{u}u + \bar{d}d - 2 \bar{s}s)$
by neglecting small mixing angle effects.
We note that the $\bar s s$ component in $\eta$ is irrelevant up to
dimension 8 if $\alpha_{S}$ corrections are ignored, since the
interpolating field (\ref{eq:nucur}) does not contain
strange quarks.
From (\ref{eq:PiOPE}) -- (\ref{eq:dGdm}),
we find that the correlation function for the $\pi NN^{*}$ coupling
vanishes in the chiral limit $m_q \to 0$, and therefore $\gpNR=0$.  
It is also found that this result remains unchanged when $\alpha_{S}$ 
corrections are included.  
In contrast,  the correlation function for the $\eta NN^*$
coupling does not vanish, and so
the coupling constant $\geNR$ remains finite.

Vanishing of the $\pslash \gamma_5$ term in the 
correlation function for $\pi NN^*$ is, in 
fact, a general consequence of chiral symmetry.  
We might have applied the soft meson theorem to the correlation function 
(\ref{eq:cor}) from the beginning.  
Using the transformation property 
\begin{equation}
\label{commutation}
[ Q_5^a , J_N] = i \gamma_5 \tau^a J_{N} \, , 
\end{equation}
we find 
\begin{eqnarray}
\Pi^{\pi^a} (p) &=& \lim_{q \to 0}  
\int d^4x e^{ipx} \bra{0} T J_N(x) \bar J_N(0) \cket{\pi^a(q)} 
\nonumber \\
&=& - \frac{i}{\sqrt{2}f_\pi} \int d^4x e^{ipx} \bra{0} [ Q_5^a , 
T J_N(x) \bar J_N(0) ] \cket{0} \nonumber \\
&=&  \frac{1}{\sqrt{2}f_\pi} 
\int d^4x e^{ipx} 
\{ \gamma_5 \tau^a, \bra{0} T J_N(x) \bar J_N(0)  \cket{0} \}  \, .
\label{Q5comm}
\end{eqnarray}
Due to the Lorentz structure 
$\bra{0}  J_N(x) \bar J_N(0)  \cket{0} \sim A \pslash + B 1$, 
the $\pslash \gamma_5$ term disappears in (\ref{Q5comm}). 

The result obtained from the commutation relation (\ref{Q5comm}) 
is more general than that in the OPE, 
since the momentum $p$ in (\ref{Q5comm}) can be arbitrary, while 
that in OPE (\ref{eq:PiOPE}) is in the deep Euclidean region.  
Therefore, Eq. (\ref{Q5comm}) implies that all transitions of 
$N^* \to \pi N$ vanish, if $N^*$ are the states that couple to $J_{N}$ 
satisfying (\ref{commutation}).  
In the real world, there are finite quark mass corrections, but as far as 
the pion sector is concerned, one may expect such transitions are strongly 
suppressed.  
Indeed, such a suppression seems to be realized phenomenologically
except for a few cases, i.e. $N(1440)$ and 
$N(1650)$~\cite{PDG}.  

To summarize shortly, our main conclusion here is that the resonance 
states which couple to the interpolating field $J_{N}$ of 
(\ref{eq:nucur}) have  
strongly suppressed couplings for $N^* \to \pi N$ in the chiral limit, 
if $J_{N}$ satisfies the commutation relation (\ref{commutation}).  
Since $J_{N}$ couples strongly to $N(1535)$ when $t \sim 
0.8$~\cite{jko}, 
the present observation is applied to this state.  
For the eta case, we do not find a relation similar to (\ref{Q5comm}), 
because the interpolating field 
$J_N$ is not a good eigen state of the U(1) axial 
charge.  
Thus we find a non-vanishing contribution to the correlation function, 
which in turn is used in the sum rule analysis 
to extract a finite coupling constant $\geNR$. 

Now, we compare the present results with those of
various low energy models.
We briefly discuss the nonrelativistic quark model, the large-$N_c$
method and effective chiral lagrangian approach.

In the nonrelativistic quark model, the negative parity nucleon is formed
by exciting one of the valence quarks into the $p$ ($l = 1$)
orbit~\cite{ik}.
Then there are two independent states for $J^P = 1/2^-$:
$\cket{1} = [l=1, S = 1/2]^{J=1/2}$ and
$\cket{2} = [l=1, S = 3/2]^{J=1/2}$,
where $S$ is the total intrinsic spin of the three quarks.
The physical state for $N(1535)$ is a linear combination of
these two states.
The coupling constants are the matrix elements of the operators,
${ \cal O}_\pi^\alpha = \sum_{i=1}^{3} \vec \sigma (i) \cdot
\vec \nabla (i) \tau^\alpha$ for the $\pi NN^*$  and
${ \cal O}_\eta = \sum_{i=1}^{3} \vec \sigma (i) \cdot
\vec \nabla (i) $ for the $\eta NN^*$.
The relative phase of the two states
$\cket{1}$ and $\cket{2}$ are then determined by the sign
of the tensor force.
In the Isgur-Karl model, it is brought by the one gluon exchange
potential, while in a more sophisticated model, there is
a significant contribution from the one pion exchange
potential also~\cite{asy}.
In both cases the phase is given such that
the interference acts destructively for $\pi NN^*$ while
constructively for $\eta NN^*$.
This explains the relatively suppressed $\gpNR$.

The suppression of $\gpNR$ is also observed
in the large-$N_c$ limit.
Assuming that the lowest baryon state develops the hedgehog
intrinsic state with $K = J + I = 0$ (here the hedgehog has negative
parity as $J^P = 1/2^-$),
it is possible to show that the matrix element for
the $\gpNR$ coupling
$\bra{N} { \cal O}_\pi^\alpha \cket{N^*}$ is of higher order in
$1/N_c$ as compared with that of $\bra{N} { \cal O}_\eta \cket{N^*}$.
This $1/N_c$ counting is in fact an example of the $I_t=J_t$ rule for
large-$N_c$ baryons~\cite{largeN}.

One may wonder if such a suppression of $\gpNR$
could be in some way related to symmetry properties.
There have been several attempts to treat positive and negative
parity baryons as chiral partners of spontaneous chiral symmetry
breaking.
DeTar and Kunihiro considered the parity doublet nucleons in the linear
sigma model of $SU(2) \times SU(2)$~\cite{DK}.
In addition to the standard chiral invariant interaction terms,
they introduced a chiral invariant mass term
between the positive and negative parity baryons.
The strength $m_{0}$ for the non-standard mass term
reduces to the mass of the would-be chiral doublet nucleons when the
chiral symmetry restores.
In the spontaneously broken phase,
the mass splitting is proportional to the non-zero expectation
value of the sigma field.
In this model, it has been shown that
$\gpNR$ is proportional to $m_{0}$  to the leading order
in $m_{0}$.
In the previous QCD sum rule study~\cite{jko}, the masses of $N$
and $N^*$ seem to get degenerate and decrease as the quark condensate
$\langle \bar q q \rangle$ is decreased.
This implies a small $m_{0}\approx 0$, which could be a possible explanation
why the coupling constant $\gpNR$ vanishes in the QCD sum rule.

Earlier, Christos investigated an effective model for
mesons and baryons by identifying the interpolating field
$B^1 \sim (qC\gamma_{5}q)q$ with the positive parity nucleon and
$B^2 \sim (qCq)q$ with the negative parity nucleon~\cite{Chr}.
Using the transformation properties of the $B^1$ and $B^2$ field, he
wrote down a chiral invariant effective Lagrangian, which leads to
the relation $\gpNR = 0$.
His lagrangian, however, does not contain the $m_{0}$ term of the
model of DeTar and Kunihiro, which is the reason for the identically
vanishing $\gpNR$.

In summary, we have studied the $\pi NN^*$ and $\eta NN^*$ coupling
constants using two point correlation functions in the soft meson 
limit.  
In the OPE, the correlation function for $\pi NN^*$ 
was calculated up to dimension eight 
and was shown to be the quantity of order ${\cal O}(m_{q})$.  
Thus the $\pi NN^*$ coupling constant is strongly suppressed.  
This suppression turns out to be a general consequence of chiral symmetry 
when the interpolating field has the suitable transformation property.  
The present results 
are consistent with the predictions of various low energy effective
approaches.
It is interesting that such a suppression of
$\gpNR$ is supported by the
chiral effective theory with parity doublet assumption.

\vspace*{0.5cm}
\noindent
{\bf Acknowledgments}\\
The authors would like to thank M. Birse for kind correspondences 
to our original manuscript and comments.  
They also thank S.H. Lee, T. Hatsuda and T. Kunihiro for 
comments and discussions.

\end{document}